\documentclass[conference]{IEEEtran}
\IEEEoverridecommandlockouts

\usepackage{cite}
\usepackage{spconf,amsmath,amssymb,amsfonts, booktabs, url}
\usepackage{algorithmic}
\usepackage{graphicx}
\usepackage{textcomp}
\usepackage{xcolor}
\def\BibTeX{{\rm B\kern-.05em{\sc i\kern-.025em b}\kern-.08em
    T\kern-.1667em\lower.7ex\hbox{E}\kern-.125emX}}
\begin{document}

\title{Frame-Wise Streaming for Efficient Learned Speech Compression
}

\name{Yuhan Liu, Jun Xu, Fengxi Zhang, Zhengxue Cheng*, Li Song*\thanks{*Corresponding authors}}
\address{School of Information Science and Electronic Engineering, Shanghai Jiao Tong University
 \\
 \{liu1025221459, xujunzz, zhangfengxi, zxcheng, song\_li\}@sjtu.edu.cn
}

\maketitle

\begin{abstract}
Speech compression is increasingly important in scenarios such as real-time and satellite communication.
Neural codecs have shown superior Rate–Distortion (RD) performance over traditional signal-processing codecs.
Models with long-range dependency encoders/decoders and entropy models outperform convolution-based structures, but they still lack streaming support since they rely on sequence-level correlations.
In this paper, we propose \textbf{StreamLSC}, a frame-level streaming learned speech compression model built on three fundamental components of streaming networks, i.e. Boundary-Cached Causal Convolution, Intra-Frame Layer Normalization, and RNN-mode RWKV.
Our model incorporates a streamable entropy quantizer module, ensuring efficient context model while preserving causal constraints.
Experiments show that the streaming model achieves performance comparable to its offline counterpart, attaining state-of-the-art rate–distortion results, significant gains in subjective evaluations, $-51.37\%$ (ViSQOL) and $-53.99\%$ (PESQ)\% BD-RATE savings on average, and improvements of 0.2154 and 0.5156 in BD-ViSQOL and BD-PESQ.
\end{abstract}


\section{Introduction}

Audio compression is fundamental for communication and storage in bandwidth-constrained environments. 
Traditional codecs such as Opus~\cite{opus} and EVS~\cite{evs} achieve high efficiency at medium bitrates, but their reliance on handcrafted features limits perceptual quality at ultra-low bitrates.

Neural codecs overcome these limitations through end-to-end training and differentiable quantization, enabling data-driven representation learning. 
Different backbone architectures have been explored: time-domain models (e.g., EnCodec~\cite{encodec}) enable low-latency reconstruction, frequency-domain models (e.g., FunCodec~\cite{funcodec}, MDCTCodec~\cite{mdctcodec}) leverage spectral representations, and attention-based designs (e.g., TS3-Codec~\cite{ts3codec}) capture long-range dependencies to enhance fidelity. 
These approaches significantly improve perceptual quality compared to traditional codecs.
\begin{figure}[tbp]
    \centering
    \includegraphics[trim=112mm 60mm -2mm 57mm, clip, width=0.85 \textwidth]{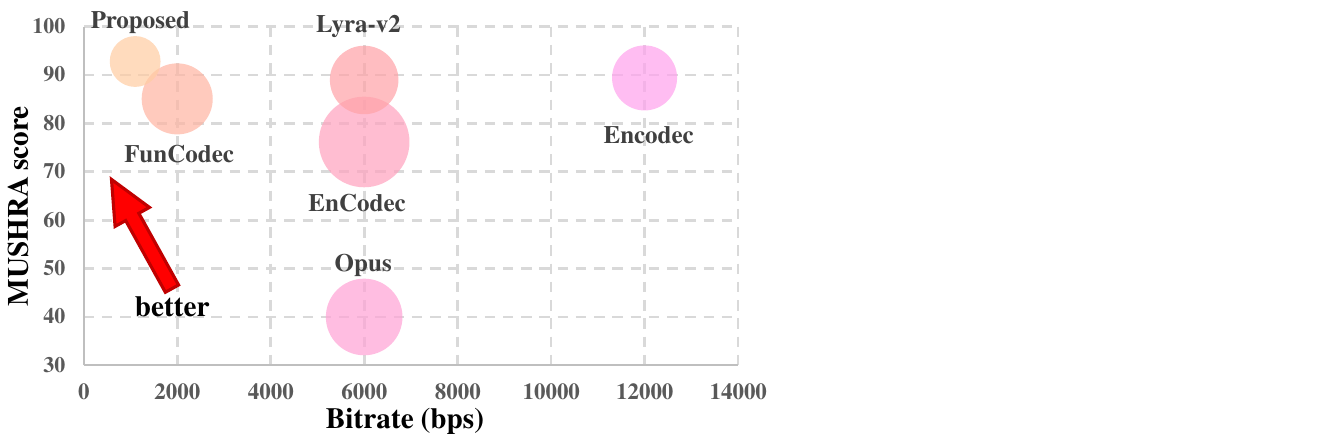}
    \vspace{-5mm}
    \caption{Subjective evaluation results using MUSHRA. The center point of each circle represents the average value, and the size reflects variability of listener ratings. Our model achieves superior perceptual quality at ultra-low bitrates.}

    \label{teaser}
\vspace{-5mm}
\end{figure}
In practical scenarios such as real-time communication and cloud-based streaming, codecs must operate in a streamable fashion, processing audio frame by frame without access to future context. 
This requirement imposes strict causality constraints on both the backbone architecture and the quantization or entropy coding modules. 
Existing streaming neural codecs primarily rely on vector-quantized representations, which naturally satisfy causality but often exhibit limited rate-distortion (RD) performance. 
Conversely, entropy-based neural compression models achieve higher coding efficiency by leveraging probabilistic latent modeling, but typically require full-sequence context and are difficult to adapt to strict streaming constraints.

These observations indicate that existing backbone designs and quantization paradigms exhibit inherent trade-offs, making it difficult to simultaneously achieve efficient RD performance and strict frame-level causality in streamable neural audio coding.

To bridge this gap, we propose a frame-level streamable codec \textbf{StreamLSC} that integrates a causal long-context backbone with a frame-level streamable entropy model, enabling efficient neural speech compression under strict streaming constraints and ultra-low bitrates.

The contributions of this paper are summarized as follows:
\begin{itemize}
    \item We propose a streamable backbone that reformulates convolution, normalization, and sequence modeling, ensuring strict causality and equivalence between streaming and offline inference.
    \item We propose a frame-level streamable entropy model, achieving efficient bit allocation under causal constraints.
    \item We demonstrate through extensive experiments that StreamLSC consistently outperforms existing baselines at ultra-low bitrates, attaining state-of-the-art RD performance and perceptual gains.
\end{itemize}


\begin{figure*}[htbp]
    \centering
    \includegraphics[trim=7mm 2mm 10mm 4mm, clip, width=1 \textwidth]{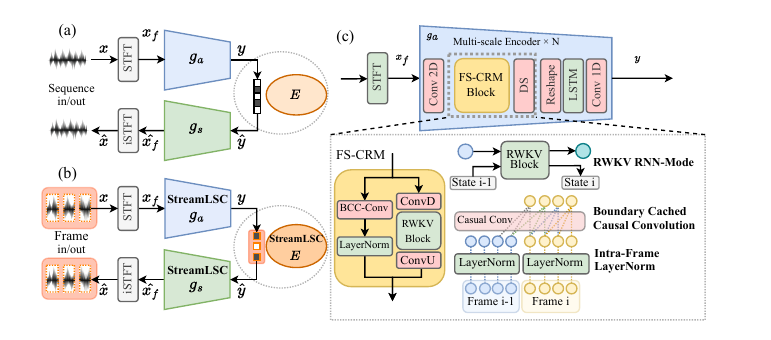}
    \vspace{-3mm}
    \caption{Illustration of model architectures. 
    (a) Previous non-streamable method. The whole audio sequence is processed at a time. 
    (b) StreamLSC, frame-level streaming processing mode. 
    (c) Streamable model components: FS-CRM block with explicit state propagation in the BCC and RWKV layers. Red blocks denote causal convolutions or convolution-based modules, while green blocks denote inherently streamable components.
}
    \label{entropy}
\vspace{-5mm}
\end{figure*}

\section{Related Work}

\subsection{Neural Speech Codecs}

Neural speech codecs employ neural networks to learn compact latent representations and achieve end-to-end optimized compression. 
Early works such as SoundStream and EnCodec adopt convolutional encoders and decoders with residual vector quantization, enabling real-time inference with subjective quality. 
Subsequent studies explore spectral-domain coding, Transformer-based latent modeling, and hierarchical temporal structures to improve long-term consistency and perceptual naturalness, highlighting the importance of long-context modeling in neural audio coding.

\subsection{Streaming Neural Speech Codecs}

Streaming neural codecs require strict causality and low latency, making architectural and quantization design particularly challenging. 
Codebook-based quantization methods, including RVQ, ERVQ~\cite{ervq}, GRVQ~\cite{grvq}, and RSVQ~\cite{streamcodec}, represent signals using multiple learned codebooks and enable flexible multi-rate operation. 
Such methods inherently preserve causality, as quantization is performed independently at each time step, and have been widely adopted in recent streaming neural codecs.

Several studies further investigate causal convolutional architectures and sliding-window attention to support real-time applications. 
However, these approaches often suffer from limited temporal context or mismatch between offline training and streaming inference, resulting in degraded RD performance and perceptual quality.

\subsection{Entropy-Based Neural Compression}

Entropy-based neural compression models estimate probability distributions of latent representations and allocate bits adaptively via entropy coding, achieving coding efficiency close to information-theoretic bounds. 
Originally developed for image compression~\cite{image_ref1,image_ref2,image_ref3,image_ref4}, these techniques have recently been extended to speech coding in our prior work~\cite{rateaware} and other studies.

In image compression, hierarchical priors, hyperpriors, and autoregressive context models significantly improve entropy estimation accuracy. 
However, applying such models to speech coding is challenging due to temporal dependencies and strict causality constraints in streaming scenarios. 
Most entropy-based speech codecs rely on full-sequence context and are unsuitable for real-time streaming applications.

\subsection{Discussion and Gap}

Existing streaming neural codecs predominantly rely on vector-quantized representations, which naturally satisfy causality but often exhibit limited RD performance. 
In contrast, entropy-based neural compression models achieve superior coding efficiency but are largely restricted to offline scenarios due to their reliance on future context.

Therefore, a fundamental gap remains: how to design a neural codec that simultaneously supports strict frame-level causality, long-context modeling, and entropy-based bit allocation. 
To address this challenge, we propose StreamLSC, which integrates a causal long-context backbone with a frame-level streamable entropy model, enabling efficient neural speech compression under strict streaming constraints.

\section{Method}

\subsection{Overview}

Fig.~\ref{entropy} illustrates our framework.
Subfigure (a) shows the prior \cite{rateaware} workflow, where the entire sequence is transformed into the frequency domain and processed into a complete bitstream.
Subfigure (b) presents our streamable workflow: audio is divided into non-overlapping frames, each frame is encoded into a partial bitstream, and explicit state propagation ensures frame-to-frame consistency. This enables real-time operation without waiting for the full sequence.

The codec consists of a streamable backbone and a streamable entropy model.
The backbone is built on the FS-CRM module, which reformulates normalization, sequence modeling, and convolution into a state-based streaming mode. Subfigure (c) shows the FS-CRM block: red denotes causal convolutions and their variants, while green denotes inherently streamable modules.
On top of this backbone, the entropy model compresses the encoder output $y$ by extracting auxiliary features $z$, which are used to predict Gaussian parameters $(\mu, \sigma)$ at the channel-group level. Conditioned on these parameters, $y$ is encoded with channel-groupwise autoregressive coding, ensuring accurate and streamable probability modeling.
This overall design guarantees causality and numerical equivalence, allowing offline training while enabling streaming deployment without parameter changes.


\subsection{Intra-Frame LayerNorm}

In our offline designs, GroupNorm is commonly used to stabilize training. However, it computes global statistics across the entire sequence, which introduces non-causal dependencies and makes it unsuitable for streaming. 
To address this, we propose \textbf{Intra-Frame LayerNorm}, where normalization is performed independently within each frame by aggregating over both feature (F) and temporal (T) dimensions:

\begin{equation}
\begin{aligned}
\mu_t &= \frac{1}{|y_t|} \sum_{i \in F,T} y_t^{(i)}, \\
\sigma_t^2 &= \frac{1}{|y_t|} \sum_{i \in F,T} (y_t^{(i)} - \mu_t)^2.
\end{aligned}
\end{equation}

This design normalizes each frame independently, without storing any state from previous frames. 
By aggregating over both frequency and temporal dimensions within the frame, it stabilizes activations, preserves short-term phonetic structure, and maintains strict causality, while avoiding the latency or complexity introduced by inter-frame caching.

\subsection{Streamable State Propagation with RWKV}

Streaming inference requires efficient state evolution under limited context rather than unrestricted global interaction.
To this end, we adopt RWKV~\cite{rwkv}, which reformulates attention as explicit state updates: at time $t$, the state $s_t$ is updated from the current input $x_t$ and the previous state $s_{t-1}$, which compactly encodes historical key-value information and residuals. 
RWKV implements the mechanism through \textit{time-mix} and \textit{channel-mix} factorization, thereby reducing complexity from quadratic to linear while maintaining long dependencies.  

A key feature of RWKV is its two numerically equivalent modes. In training, GPT-mode allows parallel full-sequence computation with CUDA acceleration, while in inference, RNN-mode enables recursive state propagation, ensuring strict causality. This dual-mode equivalence combines Transformer-level modeling capacity with RNN-like streaming efficiency, making RWKV particularly well-suited to our codec design.

\subsection{Boundary-Cached Causal Convolution}

In streaming scenarios, convolutional layers must avoid future information leakage while maintaining full receptive fields. To address this, we propose \textbf{Boundary-Cached Causal Convolution}, which preserves segment-level context using a lightweight caching mechanism.



For a causal convolution with kernel size $K$, BCC maintains the last $K-1$ time units from the previous frame as a past cache, which is concatenated with the current frame $y_t$ before convolution:
\begin{equation}
    \tilde{y}_t = \text{Conv}\big([c_{t-1}^{\text{left}}, y_t]\big).
\end{equation}
After processing, the cache is updated as
\begin{equation}
    c_t^\text{left} = \text{last}_{K-1}(y_t).
\end{equation}

The first frame is initialized using reflect padding to ensure equivalence with offline convolution, while subsequent frames rely solely on cached time units, avoiding additional padding.
This design preserves the full receptive field at frame boundaries, maintains strict causality, and ensures numerical equivalence with offline computation.
Moreover, by decoupling boundary context handling from the main convolution, BCC introduces minimal computational overhead and can be directly applied to any standard convolutional layer in a streaming pipeline.


\begin{figure*}[t!]
    \centering

    \includegraphics[trim=1mm 2mm 55mm 86mm, clip, width=\textwidth]{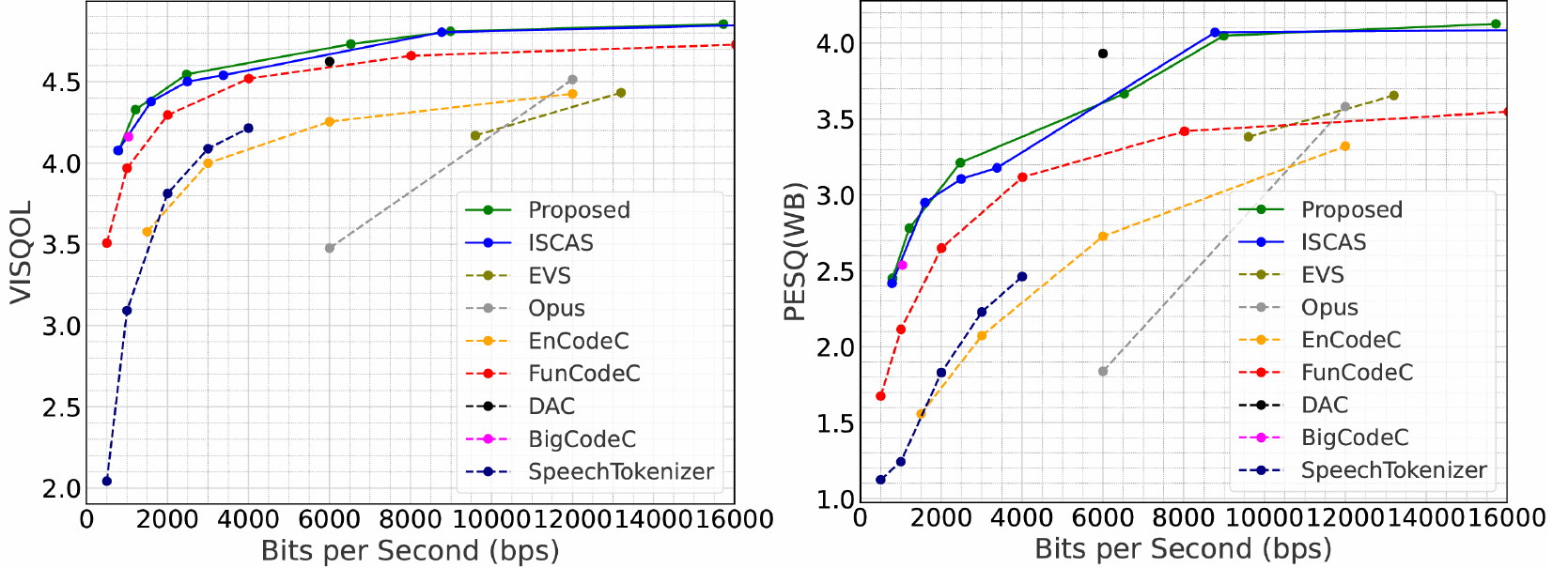}
    \vspace{-5mm}
    \caption{Objective evaluation results: ViSQOL and PESQ scores of the proposed model compared with baselines.
    The proposed StreamLSC achieves frame-level streamability while maintaining RD performance consistent with the non-streaming version, significantly outperforming the baseline solution with a performance gain of 0.2154 BD-ViSQOL and 0.5156 BD-PESQ.}
    \label{fig:rd}
    \vspace{-5mm}
\end{figure*}

\section{Experiment}

\subsection{Implementation details}

\textbf{Dataset:}
All experiments are conducted on the LibriTTS corpus~\cite{libritts}, a large-scale, high-quality speech dataset.
We follow the official train-validation-test splits to ensure fair comparison and reproducibility. 

\noindent\textbf{Loss:}
Model training is carried out using a rate–distortion objective that combines multiple loss terms.
Specifically, we include a time-domain reconstruction loss, a frequency-domain multi-scale Mel-spectrogram loss, and an adversarial loss when employing GAN training, in which discriminator feedback is incorporated to improve naturalness and reduce artifacts.  
Loss weights and hyperparameters follow the settings established in our prior work \cite{rateaware}.

\noindent\textbf{Sequential training:}
Leveraging the segment–sequence equivalence property of our design, training is performed on full sequences rather than frames.
This strategy significantly accelerates optimization while ensuring that the resulting model remains directly applicable to streaming inference. 
\noindent\textbf{Two-stage GAN training:}
Drawing inspiration from the practice in learned image compression~\cite{ganlic}, we adopt a two-stage GAN training strategy. In the first stage, the discriminator is disabled and the generator focuses solely on optimizing distortion losses. The discriminator is introduced only in the final 10 epochs for joint optimization to enhance audio naturalness. This staged approach reduces GPU memory consumption and training time during the initial phase, while achieving comparable or even slightly better model quality than standard joint training. The effectiveness of this strategy is demonstrated in the ablation study.

\noindent\textbf{Streamable lookahead}:
For inference with the frame-level streaming model, we first transform the input audio into frequency-domain features using an STFT before the encoder. 
We adopt a hop length of 160 samples. 
The first processed segment contains 1630 samples, corresponding to an 80\,ms audio frame at 16\,kHz. 
Subsequent frames consist of 1280 new samples each, with the last 350 samples of the previous frame prepended to ensure consistency in the STFT computation across frames.

\subsection{Evaluation Metrics and Baselines}

To evaluate the effectiveness of our approach, we compare it against both traditional and neural speech codecs. Conventional codecs include OPUS \cite{opus} and EVS \cite{evs}, while neural codecs include Lyra-v2 \cite{lyra}, EnCodec \cite{encodec}, FunCodec \cite{funcodec}, DAC \cite{dac}, BigCodec\cite{bigcodec}, speechtokenizer \cite{st} as well as our own non-streamable variant. 

For evaluation, we consider objective and subjective measures. Objective metrics include ViSQOL ~\cite{visqol} and PESQ ~\cite{pesq}, which assess perceptual quality and waveform fidelity. Subjective evaluation was conducted using MUSHRA~\cite{mushra} tests with 15 raters. We compared the proposed model (1.1 kbps) against FunCodec (2 kbps), EnCodec (6 kbps and 12 kbps), OPUS (6 kbps) and lyra(6 kbps). For the listening tests, we selected 12-second clips from six utterances each in the LibriTTS test-clean and test-other subsets.

\subsection{Objective Evaluation}
Fig.~\ref{fig:rd} shows the objective evaluation results. Our model achieves strong performance under both PESQ and ViSQOL evaluations. Compared to FunCodec, a non-streamable offline model, our system delivers substantial bitrate savings while improving perceptual quality: BD-RATE reductions of $-51.37\%$ (ViSQOL) and $-53.99\%$ (PESQ) indicate nearly half the bitrate is needed to achieve the same quality. At equivalent bitrates, BD-ViSQOL gains of $+0.2154$ and BD-PESQ of $+0.5156$ further confirm consistent improvements in perceptual fidelity. Our model also demonstrates superior metrics compared to DAC, and achieves competitive performance against BigCodec (159M parameters) even at an ultra-low bitrate of 1000 bps. SpeechTokenizer shows inferior performance on these two metrics, as it is primarily designed for speech language models, where the first RVQ layer encodes semantic-level representations and does not adequately preserve fine-grained acoustic details.

These results align with the RD curves, where our codec consistently dominates FunCodec across all operating points. Notably, although FunCodec and BigCodec are offline models with access to full-sequence context, our frame-level streaming model achieves comparable or superior performance, highlighting the effectiveness of the proposed causal long-context backbone and streamable entropy model. This demonstrates that strict causality constraints do not necessarily compromise coding efficiency when properly modeled.

The improvements can be attributed to our stronger encoder–decoder design together with an advanced entropy model that combines hyperprior information with channel-group autoregressive coding, which reduces redundancy within the latent codes, leading to higher coding efficiency and better preservation of fine-grained perceptual details, particularly at low bitrates.

\subsection{Subjective Evaluation}

Fig.\ref{teaser} presents subjective MUSHRA results.
Our codec achieves an average score of 92.8, close to the reference (95.3) and consistently higher than all baselines. Despite operating at only 1.1 kbps, it surpasses FunCodec (2 kbps), EnCodec and Lyra-v2 at 6 kbps, and even EnCodec at 12 kbps, while outperforming OPUS (40.0). Moreover, it attains a median of 100 with a narrow interquartile range [89, 100], indicating greater stability compared to FunCodec and EnCodec. Audio samples used in the subjective evaluation are available at \url{https://buhaneryu.github.io/Voice_Demo_Website/}. Combined with the objective BD-RATE analysis, these results confirm that StreamLSC delivers superior perceptual quality and coding efficiency at ultra-low bitrates.
\subsection{Ablation Study}

\begin{table}[t]
\centering
\caption{Effect of frame length on quality (ViSQOL).}
\vspace{2mm}
\begin{tabular}{lcccc}
\toprule
Frame length & 80\,ms & 160\,ms & 320\,ms & 640\,ms \\
\midrule
ViSQOL  & 4.8257 & 4.8283 & 4.8315 & 4.8319 \\
\bottomrule
\end{tabular}
\label{tab:ablation_frame}
\vspace{-2mm}
\end{table}

We conducted an ablation study on frame length. The results in Table~\ref{tab:ablation_frame} indicate that the model achieves comparable performance across all tested frame lengths. 
This demonstrates that our model can maintain high reconstruction quality even with the shortest 80\,ms frames, highlighting its effectiveness for low-latency streaming scenarios.

\begin{table}[t]
\centering

\caption{Comparison of training strategies on test\_clean and test\_others datasets.}
\vspace{3mm}
\begin{tabular}{lccc}
\toprule
Training Strategy & Bitrate (bps) & ViSQOL & PESQ \\
\midrule
Full GAN Training & 3529.00 & 4.4500 & 3.1800 \\
Two-stage GAN Training & 3377.29 & 4.5398 & 3.1780 \\
\bottomrule
\end{tabular}
\label{tab:ablation_training}
\vspace{-2mm}
\end{table}

Additionally, we performed an ablation study on the training methodology. As shown in Table~\ref{tab:ablation_training}, we compared two training strategies: (1) training with discriminator and generator throughout the process, and (2) our proposed two-stage GAN training approach. The results demonstrate that the two-stage training achieves a lower bitrate (3377.29 bps vs. 3529 bps) while maintaining comparable PESQ scores (3.178 vs. 3.180) and superior ViSQOL scores (4.54 vs. 4.45). 

This improvement may be attributed to the fact that in the initial stage without the discriminator, the codec can focus solely on distortion optimization without the adversarial loss interfering with the reconstruction objective. Furthermore, the absence of the discriminator reduces GPU memory consumption, enabling the use of larger batch sizes for more stable training dynamics and better convergence properties.

\subsection{Runtime Analysis}

\begin{table}[t]
\centering
\caption{Model parameters and real-time factor (RTF) on RTX 4090.}
\vspace{2mm}
\begin{tabular}{lcc}
\toprule
Model  & RTF \\
\midrule
StreamLSC (Ours)   & $5.39\times 10^{-2}$ \\
FunCodec           & $5.45\times 10^{-3}$ \\
EnCodec            & $4.695\times 10^{-3}$ \\
DAC                & $6.367\times 10^{-3}$ \\
BigCodec          & $1.504\times 10^{-2}$ \\
\bottomrule
\end{tabular}
\label{tab:inference_rtf}
\vspace{-5mm}
\end{table}

We evaluate runtime on an RTX 4090 GPU, where RTF is defined as the ratio between inference time and audio duration.
Our model contains 149.2M parameters and achieves an RTF of 0.0539, which indicates that it is capable of operating in real-time scenarios.
This demonstrates that the proposed scheme provides a favorable balance between model complexity and reconstruction efficiency, making it suitable for practical deployment.

\section{Conclusion}

We propose a frame-level streamable backbone and entropy model with streaming components (Boundary-Cached Causal Convolution, Intra-frame LayerNorm, and RNN-mode RWKV) to construct the streamable learned speech compression.
These components capture both local and non-local correlations while satisfying causal constraints, delivering excellent RD performance for the overall compression scheme.
Experimental results demonstrate that the proposed frame-level streaming learned speech compression achieves state-of-the-art RD performance, with significant gains in subjective evaluations and improvements of $-51.37\%$ (ViSQOL) and $-53.99\%$ (PESQ) in BD-RATE, as well as 0.2154 and 0.5156 in BD-ViSQOL and BD-PESQ on objective metrics.

\section{Acknowledgement}

This work was partly supported by the NSFC(62431015, 62571317, 62501387), the Fundamental Research Funds for the Central Universities, Shanghai Key Laboratory of Digital Media Processing and Transmission under Grant 22DZ2229005, 111 project BP0719010.


\begin{thebibliography}{00}
\bibitem{opus} J.-M. Valin, K. Vos, and T. Terriberry, ``Definition of the opus audio codec,'' 2012.

\bibitem{evs} M. Dietz et al., ``Overview of the EVS codec architecture,'' in \emph{Proc. 2015 IEEE Int. Conf. Acoustics, Speech and Signal Processing (ICASSP)}, pp. 5698--5702, 2015.

\bibitem{lyra} N. Zeghidour, A. Luebs, A. Omran, J. Skoglund, and M. Tagliasacchi, ``Soundstream: An end-to-end neural audio codec,'' IEEE/ACM Trans. Audio, Speech, and Language Process., vol. 30, pp. 495--507, 2021.
\bibitem{encodec} A. D{\'e}fossez, J. Copet, G. Synnaeve, and Y. Adi, ``High fidelity neural audio compression,'' arXiv:2210.13438, 2022. [Online]. Available: https://arxiv.org/abs/2210.13438
\bibitem{streamcodec} X.-H. Jiang, Y. Ai, R.-C. Zheng, and Z.-H. Ling, ``A streamable neural audio codec with residual scalar-vector quantization for real-time communication,'' IEEE Signal Process. Lett., vol. 32, pp. 1645--1649, 2025.
\bibitem{rateaware} J. Xu, Z. Cheng, G. Chi, Y. Liu, Y. Hu, and L. Song, ``Rate-aware learned speech compression,'' in \emph{Proc. 2025 IEEE Int. Symp. Circuits and Systems (ISCAS)}, pp. 1--5, 2025.


\bibitem{ervq} R.-C. Zheng, H.-P. Du, X.-H. Jiang, Y. Ai, and Z.-H. Ling, ``ERVQ: Enhanced residual vector quantization with intra-and-inter-codebook optimization for neural audio codecs,'' IEEE Trans. Audio, Speech and Language Process., 2025.
\bibitem{grvq} D. Yang, S. Liu, R. Huang, J. Tian, C. Weng, and Y. Zou, ``Hifi-codec: Group-residual vector quantization for high fidelity audio codec,'' arXiv:2305.02765, 2023. [Online]. Available: https://arxiv.org/abs/2305.02765

\bibitem{funcodec} Z. Du, S. Zhang, K. Hu, and S. Zheng, ``Funcodec: A fundamental, reproducible and integrable open-source toolkit for neural speech codec,'' in \emph{Proc. ICASSP 2024-2024 IEEE Int. Conf. Acoustics, Speech and Signal Processing (ICASSP)}, pp. 591--595, 2024.
\bibitem{mdctcodec} X.-H. Jiang, Y. Ai, R.-C. Zheng, H.-P. Du, Y.-X. Lu, and Z.-H. Ling, ``Mdctcodec: A lightweight mdct-based neural audio codec towards high sampling rate and low bitrate scenarios,'' in \emph{Proc. 2024 IEEE Spoken Language Technology Workshop (SLT)}, pp. 540--547, 2024.
\bibitem{ts3codec} H. Wu, N. Kanda, S. E. Eskimez, and J. Li, ``Ts3-codec: Transformer-based simple streaming single codec,'' arXiv:2411.18803, 2024. [Online]. Available: https://arxiv.org/abs/2411.18803
\bibitem{visqol} M. Chinen, F. S. C. Lim, J. Skoglund, N. Gureev, F. O'Gorman, and A. Hines, ``ViSQOL v3: An open source production ready objective speech and audio metric,'' in \emph{Proc. 2020 Twelfth Int. Conf. Quality of Multimedia Experience (QoMEX)}, pp. 1--6, 2020.
\bibitem{pesq} A. W. Rix, J. G. Beerends, M. P. Hollier, and A. P. Hekstra, ``Perceptual evaluation of speech quality (PESQ)-a new method for speech quality assessment of telephone networks and codecs,'' in \emph{Proc. 2001 IEEE Int. Conf. Acoustics, Speech, and Signal Processing}, vol. 2, pp. 749--752, 2001.
\bibitem{mushra} ITU-R, ``Method for the subjective assessment of intermediate quality level of audio systems,'' Int. Telecommun. Union, Tech. Rep. BS.1534-3, 2015.
\bibitem{libritts} H. Zen et al., ``Libritts: A corpus derived from librispeech for text-to-speech,'' arXiv:1904.02882, 2019. [Online]. Available: https://arxiv.org/abs/1904.02882
\bibitem{rwkv} B. Peng et al., ``Eagle and finch: Rwkv with matrix-valued states and dynamic recurrence,'' arXiv:2404.05892, 2024. [Online]. Available: https://arxiv.org/abs/2404.05892
\bibitem{image_ref1} J. Ball{\'e}, D. Minnen, S. Singh, S. J. Hwang, and N. Johnston, ``Variational image compression with a scale hyperprior,'' arXiv:1802.01436, 2018. [Online]. Available: https://arxiv.org/abs/1802.01436
\bibitem{image_ref2} Z. Cheng, H. Sun, M. Takeuchi, and J. Katto, ``Learned image compression with discretized gaussian mixture likelihoods and attention modules,'' in \emph{Proc. IEEE/CVF Conf. Computer Vision and Pattern Recognition}, pp. 7939--7948, 2020.
\bibitem{image_ref3} D. Minnen and S. Singh, ``Channel-wise autoregressive entropy models for learned image compression,'' in \emph{Proc. 2020 IEEE Int. Conf. Image Processing (ICIP)}, pp. 3339--3343, 2020.
\bibitem{image_ref4} J. Liu, H. Sun, and J. Katto, ``Learned image compression with mixed transformer-cnn architectures,'' in \emph{Proc. IEEE/CVF Conf. Computer Vision and Pattern Recognition}, pp. 14388--14397, 2023.
\bibitem{bigcodec} D. Xin, X. Tan, S. Takamichi, and H. Saruwatari, ``Bigcodec: Pushing the limits of low-bitrate neural speech codec,'' arXiv:2409.05377, 2024. [Online]. Available: https://arxiv.org/abs/2409.05377
\bibitem{dac} R. Kumar, P. Seetharaman, A. Luebs, I. Kumar, and K. Kumar, ``High-fidelity audio compression with improved rvqgan,'' in \emph{Advances in Neural Information Processing Systems}, vol. 36, pp. 27980--27993, 2023.
\bibitem{ganlic} S. Iwai, T. Miyazaki, and S. Omachi, ``Controlling rate, distortion, and realism: Towards a single comprehensive neural image compression model,'' in \emph{Proc. IEEE/CVF Winter Conf. Applications of Computer Vision (WACV)}, 2024.
\bibitem{st} X. Zhang, D. Zhang, S. Li, Y. Zhou, and X. Qiu, ``SpeechTokenizer: Unified speech tokenizer for speech large language models,'' \emph{arXiv:2308.16692}, 2023.


\end{thebibliography}
\end{document}